\documentclass[%
 reprint,
 amsmath,amssymb,
 aps,
]{revtex4-1}

\usepackage{graphicx}
\usepackage{dcolumn}
\usepackage{bm}
\usepackage{xspace}
\usepackage[overload]{textcase}

\usepackage[ansinew]{inputenc}

\newcommand{\pt}{\ensuremath{p_{\rm T}}\xspace}

\newcommand{\nch}{\ensuremath{N_{\rm ch}}\xspace}
\newcommand{\nmpi}{\ensuremath{{\rm N}_{\rm mpi}}\xspace}

\begin{document}

\preprint{APS/123-QED}

\title{Multi-Parton Interactions in pp collisions from Machine Learning-based regression}

\author{Antonio Ortiz}
 \email{antonio.ortiz@nucleares.unam.mx}
\author{Antonio Paz}  
\author{Jos\'e D. Romo} 
\author{Sushanta Tripathy}%
\email{sushanta.tripathy@cern.ch}
\author{Erik A. Zepeda}%
\affiliation{%
Instituto de Ciencias Nucleares, Universidad Nacional Aut\'onoma de M\'exico,\\
 Apartado Postal 70-543, M\'exico Distrito Federal 04510, M\'exico 
}%

\author{Irais Bautista}
\affiliation{%
Facultad de Ciencias Físico Matem\'aticas, Benemérita Universidad Aut\'onoma de Puebla, 1152,  Puebla  72570, M\'exico; \\
Departamento de F\'isica,
Centro de Investigación y de Estudios Avanzados del Instituto Polit\'ecnico Nacional, 14-740, M\'exico Distrito Federal  07000, M\'exico
}%

\date{\today}

\begin{abstract}
Multi-Parton Interactions (MPI) in pp collisions have attracted the attention of the heavy-ion community since they can help to elucidate the origin of collective-like effects discovered in small collision systems at the LHC. In this work, we report that in PYTHIA~8.244, the charged-particle production in events with a large number of MPI (${\rm N}_{\rm mpi}$) normalized to that obtained in minimum-bias pp collisions shows interesting features. After the normalization to the corresponding $\langle {\rm N}_{\rm mpi} \rangle$, the ratios as a function of $p_{\rm T}$ exhibit a bump at $\pt\approx3$\,GeV/$c$; and for higher $p_{\rm T}$ ($>8$\,GeV/$c$), the ratios are independent of ${\rm N}_{\rm mpi}$. While the size of the bump increases with increasing ${\rm N}_{\rm mpi}$, the behavior at high $p_{\rm T}$ is expected from the ``binary scaling'' (parton-parton interactions), which holds given the absence of any parton-energy loss mechanism in PYTHIA. The bump at intermediate $p_{\rm T}$ is reminiscent of the Cronin effect observed for the nuclear modification factor in p--Pb collisions. In order to unveil these effects in data, we propose a strategy to construct an event classifier sensitive to MPI using Machine Learning-based regression. The study is conducted using TMVA, and the regression is performed with Boosted Decision Trees (BDT). Event properties like forward charged-particle multiplicity, transverse spherocity and the average transverse momentum ($\langle p_{\rm T} \rangle$) are used for training. The kinematic cuts are defined in accordance with the ALICE detector capabilities. For the validation of the method and to find possible model dependence, we also compare the results from PYTHIA~8.244 with HERWIG~7.1. In addition, we also report that if we apply the trained BDT on existing (${\rm INEL}>0$) pp data, i.e. events with at least one primary charged-particle within $|\eta|<1$,  the average number of MPI in pp collisions at $\sqrt{s}=5.02$ and 13\,TeV are 3.76$\pm1.01$ and 4.65$\pm1.01$, respectively.

\end{abstract}

\maketitle


\section{Introduction}

The goal of the heavy-ion program is to understand the behavior of Quantum Chromo-Dynamics (QCD) at high temperatures and densities. Results at the Large Hadron Collider (LHC) confirmed the formation of a new form of matter characterized by deconfinement, which is compatible with the theoretically predicted Quark-Gluon Plasma (QGP)~\cite{Busza:2018rrf}. The main conclusions arose from comparisons of heavy-ion data with reference data, such as minimum-bias pp and p--A collisions, where no signatures of  jet quenching were observed. Surprisingly, the multiplicity-dependent analysis of the pp data at $\sqrt{s} = 7$\,TeV from the LHC, unveiled very similar azimuthal anisotropies as in heavy-ion collisions~\cite{Khachatryan:2010gv}. The analysis was further extended to lower and higher energies~\cite{Khachatryan:2016txc}, as well as for other systems such as p--Pb collisions at $\sqrt{s_{\rm NN}}=5.02$\,TeV~\cite{Abelev:2012ola,Aaij:2015qcq}. Moreover, reports on the enhancement of (multi-)strange hadrons in pp and p--Pb collisions~\cite{ALICE:2017jyt,Adam:2015vsf}, as well as the mass ordering in the hadron \pt spectra~\cite{Acharya:2018orn,Adam:2016dau} suggest that collective phenomena are present at the LHC energies even in small systems. 

Naturally,  it  is  suggested  that  the new phenomena could have the same origin as in heavy-ion collisions, namely, the hydrodynamic response  of  the  produced  medium  to  the  initial  shape  of the interaction region in the transverse plane~\cite{Bozek:2011if}. However, the main concern relies on the applicability of hydrodynamics to small non-equilibrium systems. On the other hand, from the initial state perspective the azimuthal anisotropy is due to the presence of initial state correlations in the nuclear wave functions of the incoming nuclei~\cite{Schlichting:2016sqo}. The main concern is whether azimuthal anisotropies established during the initial stages of the collision can survive subsequent final state interactions~\cite{Strickland:2018exs}. Another approach relies on partonic and hadronic transport models, for example AMPT~\cite{Lin:2004en}. This model qualitatively, and sometimes quantitatively, describes small system flow signals for various collision systems and energies. The big issue is that in contrast to fluid dynamic simulation, its applicability relies on a sufficiently long mean free path, which is hard to reconcile with the idea of the strongly coupled hydrodynamic system.

Other alternative microscopic descriptions, for example, models based on QCD theory of Multi-Parton Interactions (MPI) are proposed to explain collectivity from interference effects in hadronic collisions with \nmpi parton-parton scatterings  (or ``sources'')~\cite{Blok:2017pui,Blok:2018xes}.  Another example is PYTHIA~\cite{Sjostrand:2014zea}, which uses the string fragmentation including interactions between strings~\cite{Bierlich:2017vhg} along with an initial state provided by a smooth distribution of MPI. In particular, particle production as a function of \nmpi unveils collective-like effects in PYTHIA~8 simulations with color reconnection (CR)~\cite{Ortiz:2013yxa}. Results within the string percolation framework have also been reported~\cite{Bautista:2015kwa}. Moreover, HERWIG~7.1~\cite{Bellm:2015jjp}, which recently updated its CR model, has significantly improved the description of hadron-to-pion ratios as a function of charged-particle multiplicity~\cite{Acharya:2020zji}.  From the above discussion, it is clear that the unified description of the observed phenomena across different  collision systems is still an open problem~\cite{Nagle:2018nvi}.  

From the experimental side, one challenge for pp collisions is the strong correlation between multiplicity (sensitive to low-\pt particles) and hard physics (high-\pt particles)~\cite{Acharya:2019mzb,Ortiz:2016kpz}. It has been shown that the correlation is reduced if the event multiplicity is determined in a pseudorapidity region far from where the observable of interest is measured. However, an additional treatment of the unwanted particle correlations (originated e.g. from jets) has to be implemented in data analysis. Having an event-activity estimator with less selection bias could help to improve the comparison of pp collisions with larger systems like those created in p--A and A--A collisions.  This motivates the introduction of different multiplicity estimators, for instance, the relative transverse activity classifier which aims at studying the hadronization in events with an extreme underlying event~\cite{Martin:2016igp}. However, this requires a cut on the transverse momentum of the leading particle, which biases the sample towards hard processes in a nontrivial way~\cite{Ortiz:2018vgc,Ortiz:2017jaz,Ortiz:2020dph}. In this paper, we propose the use of a Machine Learning-based regression to build a more inclusive event classifier aimed at reducing the selection biases and increasing its sensitivity to MPI.  This event classifier can help to test ideas like collectivity from interference~\cite{Blok:2017pui,Blok:2018xes} or CR~\cite{Ortiz:2013yxa}. Based on this approach, we also estimate $\langle\nmpi\rangle$ for the existing so-called ${\rm INEL}>0$ ALICE data~\cite{Acharya:2019mzb}. 

The paper is organised as follows:  section 2 describes the multivariate analysis, where the input variables and the models used for the study are discussed.  Results are presented in section 3, and finally section 4 contains a summary and outlook.

\section{Multivariate MPI-activity estimation}
Our approach relies on a multivariate regression technique based on Boosted Decision Trees (BDT) with gradient boosting training~\footnote{ In our study we consider the following parameters: NTrees=2000, Shrinkage=0.1BaggedSampleFraction=0.5, nCuts=20, MaxDepth=4 (details can be found in Ref.~\cite{hoecker2007tmva})}. This is done using the Toolkit for Multivariate Analysis (TMVA) framework which provides a ROOT-integrated machine learning environment for the processing and parallel evaluation of multivariate classification and regression technique~\cite{hoecker2007tmva}. In particular, the construction of an event classifier sensitive to the MPI activity (\nmpi) can be considered as a regression problem where a given set of input variables tries to minimize the loss function. Such a loss function describes how the model is predictive with respect to the training data. For the regression problem, TMVA implements the Huber loss function~\cite{huber1964}.

The training for the MPI-activity estimation is performed on simulated samples  of pp collisions at 13\,TeV. PYTHIA~8.244~\cite{Sjostrand:2014zea} event generator (tune 4C~\cite{Corke:2010yf}) is used in our studies. Two samples are employed to check the performance of the method for the estimation of the average \nmpi both in MB and high \nmpi events. The first sample yields a flat \nmpi distribution and the second one is that obtained from MB events.   The goal of the analysis is to build an event classifier sensitive to \nmpi, therefore \nmpi is our target variable. The MVA uses several input variables, which are chosen given their correlation with \nmpi. Another important factor related to the choice of the variables relies on how well  PYTHIA~8.244 describes such features of data. We choose PYTHIA~8.244 tune 4C as it has been tuned to describe the early LHC data (pp collisions at $\sqrt{s} = 7$\,TeV). Given that the tune only considers observables with unidentified primary-charged particles, only those particles are used in the present analysis to train the BDT. These observables are listed below:

\begin{itemize}
  \item {\bf Transverse spherocity:}  this quantity allows one to know whether a dijet-like structure is present in the event~\cite{Ortiz:2017jho}. It is defined for a unit vector $\mathbf{ \hat{\rm \mathbf{n}}_{\rm \mathbf{s}} }$ which minimizes the ratio:

\begin{equation}
S_{\rm 0} \equiv \frac{\pi^{2}}{4}  \underset{\bf \hat{n}_{\rm \bf{s}}}{\text{min}}  \left( \frac{\sum_{i}|{\vec p}_{{\rm T},i} \times { \bf \hat{n}_{\rm \bf{s}} }|}{\sum_{i}p_{{\rm T},i}}  \right)^{2},
\end{equation}
where the sum runs over all primary charged particles with $\pt>0.15$\,GeV/$c$ and within $|\eta|<0.8$. In agreement with ALICE requirements~\cite{Acharya:2019mzb}, only events with more than two particles are selected. As outlined in Ref.~\cite{Acharya:2019mzb}, spherocity has  some important features:
\begin{itemize}
\item The vector products are linear in particle momenta, therefore spherocity is a collinear safe quantity in pQCD.
\item The lower limit of spherocity ($S_{0}\rightarrow0$) corresponds to event topologies where all transverse momentum vectors are (anti)parallel or the sum of the \pt is dominated by a single track.
\item The upper limit of spherocity ($S_{0}\rightarrow1$) corresponds to event topologies where transverse momentum vectors are ``isotropically'' distributed. $S_{0}=1$ can only be reached in the limit of an infinite amount of particles.
\end{itemize}

\item {\bf Average transverse momentum:} the first moment of the charged-particle transverse momentum spectrum and its correlation with the charged particle multiplicity, encodes information about the underlying particle production mechanism. In particular, in PYTHIA the rise of the average \pt with the event multiplicity can only be explained if collective-like effects are included in the simulations (color reconnection). Therefore, this quantity is sensitive to the hadronization mechanism. In this analysis, the average \pt is obtained event-by-event considering charged particles with transverse momentum above 0.15\,GeV/$c$ and within $|\eta|<0.8$.

\item{\bf Forward multiplicity:}  it is determined within the pseudorapidity regions $2.8<\eta<5.1$ and $-3.7<\eta<-1.7$, which matches the intervals covered by the ALICE VZERO detector. This has been used by the experiment in order to reduce the autocorrelations, which may affect the spectral shape of the transverse momentum distribution.

\end{itemize}

The method was trained using simulations (tune 4C) at the highest center-of-mass energy achieved by the LHC during run II (13\,TeV).  For training, different conditions were varied to estimate a systematic uncertainty on the target variable. 
\begin{itemize}
\item For training,  average \pt and mid-pseudorapidity multiplicity were used instead the set listed above.
\item Extreme cases of \nmpi distributions were considered. We compared the results obtained using the \nmpi distribution provided by the tune 4C with those where a flat \nmpi distribution was assumed in the simulations.  
\item To check the robustness of the trained BDT, the model dependence was evaluated through a variation of the PYTHIA tune. The tunes 2C~\cite{Corke:2010yf} and Monash 2013~\cite{Skands:2014pea}, which give the worst and best description of the LHC data, respectively, were used for training instead tune 4C. 
\end{itemize}
The trained method was further tested on pp collisions  at different center-of-mass energies simulated with the tune 4C. In addition, the target variable ($\nmpi^{\rm reg.}$) was verified to satisfy the known behavior of \nmpi: it should be independent of CR, and for simulations without MPI it should be consistent to one. Therefore, the trained BDT were also applied to simulations which do not incorporate either MPI or CR. It is worth mentioning that simulations without color reconnection also allow for the study of the robustness of the trained BDT against collective-like effects~\cite{Ortiz:2013yxa}. The variations of the target value with respect to the real number of MPI was assigned as systematic uncertainty. 

\begin{figure}[b]
\includegraphics[width=0.5\textwidth]{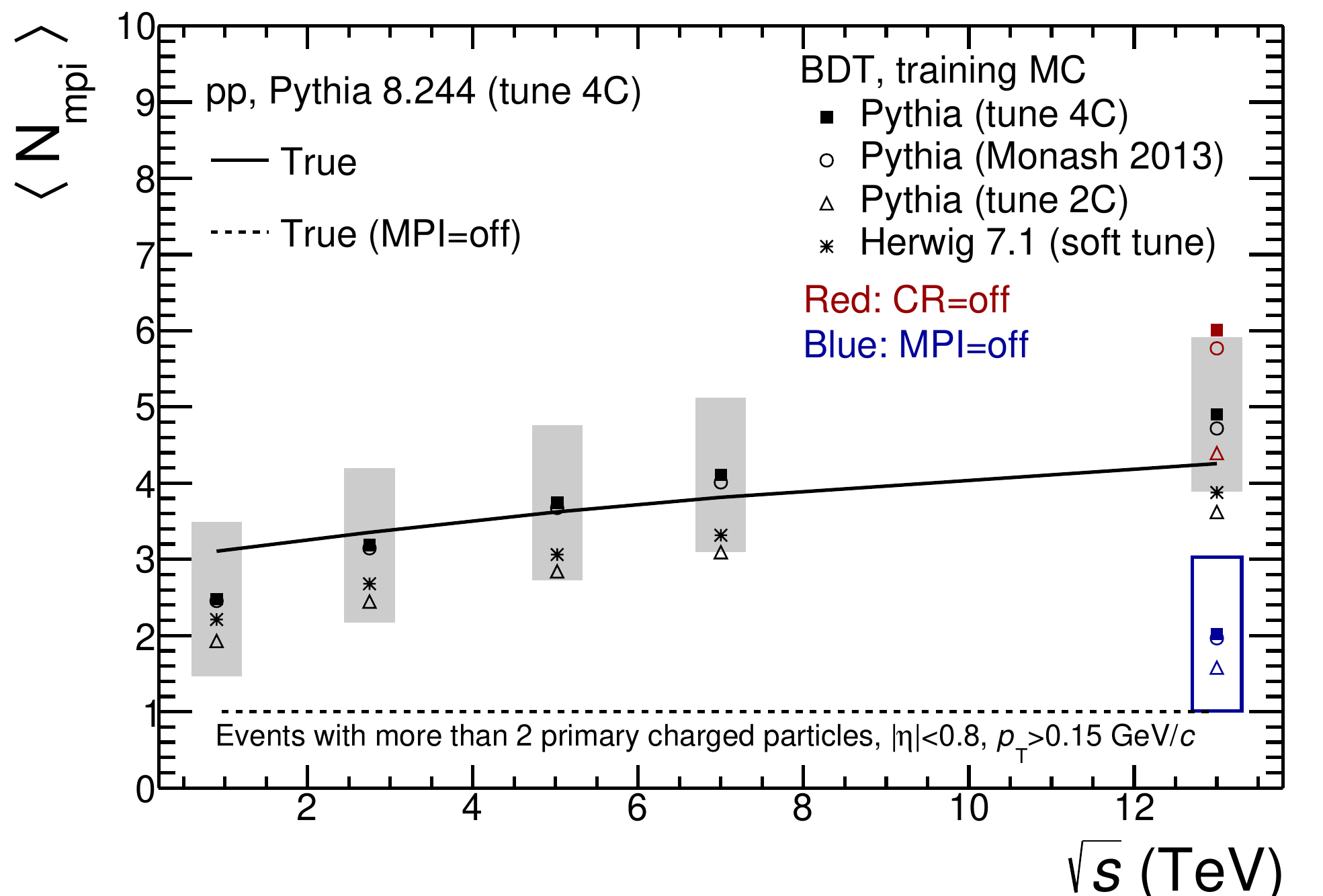}
\caption{\label{fig:1} Average number of MPI as a function of the center-of-mass energy for minimum-bias pp collisions simulated with PYTHIA~8.244 tune 4C (solid line). The dotted line represents the expected value for simulations when MPI is switched off. The trained BDT were applied to tune 4C simulations and the obtained values are displayed for the cases which use the PYTHIA tunes: 4C (full markers), Monash 2013 (open circles) and 2C (open triangles) for training. Results which uses HERWIG~7 simulations for training are also displayed (stars). The boxes around the markers correspond to the systematic uncertainty associated to our reference results. Results for simulations which do not include color reconnection (red) or MPI (blue) are also shown for $\sqrt{s}=13$\,TeV.}
\end{figure}

\begin{figure}[b]
\includegraphics[width=0.5\textwidth]{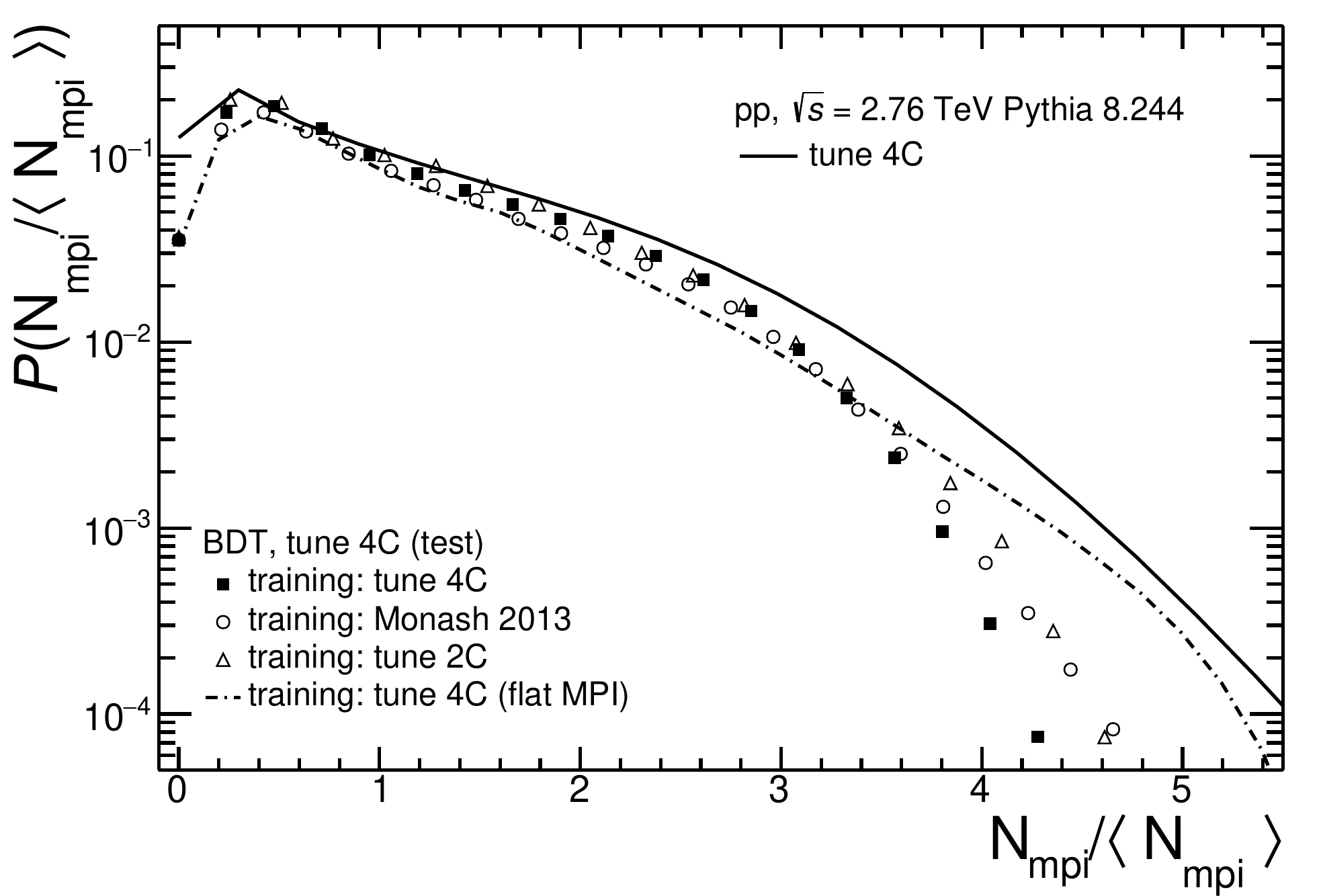}
\caption{\label{fig:1b} Distribution of the self-normalized \nmpi for minimum-bias pp collisions at $\sqrt{s}=2.76$\,TeV simulated with PYTHIA~8.244 tune 4C (solid line).  The distributions obtained from regression are displayed for the cases which use the tunes: 4C (full markers), Monash 2013 (open circles) and 2C (open triangles) for the training of BDT. BDT were also trained using simulations with a flat \nmpi distribution, the result from regression is also shown (dashed line).}
\end{figure}

\section{Results}

Given that no data on \pt spectra as a function of spherocity are available~\cite{Acharya:2019mzb}, for minimum bias pp collisions, the reference results are obtained considering only average \pt and mid-pseudorapidity multiplicity as input variables. This allows us to analyze the existing data using the trained BDT. Figure~\ref{fig:1} illustrates the performance of regression for minimum-bias simulations at different center-of-mass energies ($\sqrt{s}=0.9$, 2.76, 5.02, 7, and 13 TeV), the boxes around the points correspond to the sigma of the $\nmpi^{\rm reg}-\nmpi$ distribution. Within uncertainties, the method reproduces the expected behavior for $\langle\nmpi^{\rm reg.}\rangle(\sqrt{s})$,  albeit the modest energy dependence of \nmpi. This feature is crucial for event-by-event classification in terms of the MPI activity. The performance of the method in simulations without color reconnection is also shown. Within uncertainties, $\langle\nmpi^{\rm reg}\rangle$ is independent of color reconnection, suggesting that the method is robust against collective-like effects. Moreover, since in this case MPI produce particles nearly independently of each other, the average transverse momentum is independent of \nmpi, and hence of the charged-particle multiplicity.  In other words, simulations without CR significantly overestimate (underestimate) the measured multiplicity of the events (the average \pt of the final hadrons). However, even under these challenging conditions, the method preserves its sensitivity to MPI.  Last but not least, when MPI are not activated in the simulations, the method gives a value that within one sigma is consistent with unity. 

\begin{figure*}
\includegraphics[width=0.99\textwidth]{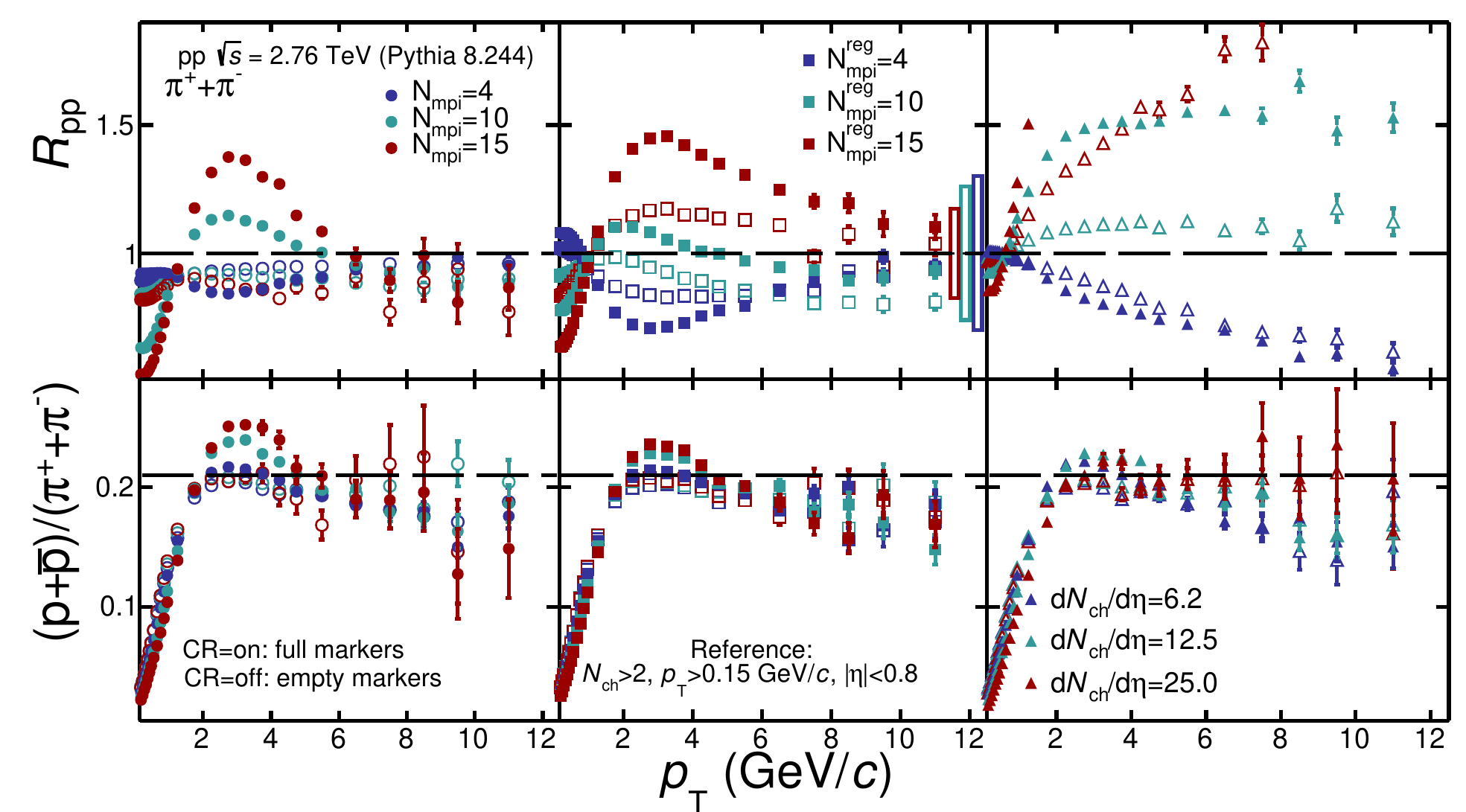}
\caption{\label{fig:2} Primary charged pion $R_{\rm pp}$ as a function of \pt (top) and proton-to-pion ratio as a function of \pt (bottom). Results are presented for different event classes based on the actual number of multi-parton interactions (left), the target variable (middle), and mid-pseudorapidity charged-particle multiplicity (right). Results from simulations including color reconnection are shown with full markers, while the case where color reconnection is switched off is displayed with empty markers. The boxes around one indicate the estimated uncertainty associated to event selection.}
\end{figure*}

Figure~\ref{fig:1} also displays $\langle\nmpi^{\rm reg.}\rangle$ as a function of $\sqrt{s}$ for pp collisions (tune 4C) under the variation of the PYTHIA tune used for training.  Within one sigma, the results are independent of the tune used for training. The largest deviation with respect to the reference results (tune 4C for training) is observed for the tune 2C. This parametrization was tuned to TEVATRON data, and it was not designed to be  a ``complete'' tuning of the generator~\cite{Corke:2010yf}. It was provided, instead, as a starting point for more sophisticated tunes using the LHC data. The model 4C, on the other hand, used the early LHC MB and UE data (pp at $\sqrt{s}=0.9$ and 7\,TeV) for tuning, which the model 2C significantly underestimated. The Monash 2013 model is tuned to a bigger set of LHC data~\cite{Skands:2014pea}. Contrary to the other tunes, Monash starts from a more careful tune to LEP data, and it involves several parameter changes. Given the important differences among these parametrizations, our results suggest that within one sigma, the proposed method is robust against variations in the \nmpi model.  And therefore, it can be used to learn about the initial state in pp collisions given by the number of partonic scatterings. 

The effects of the hadronization model used in the Monte Carlo generator were further investigated using HERWIG~7.1 simulations~\cite{Bellm:2019zci}. In the latest version of HERWIG, the modelling of underlying event is based on the eikonal approach, where the MPI activity is modeled as additional semi-hard and soft partonic interactions. Figure~\ref{fig:1} also displays the regression value obtained when pp collisions at $\sqrt{s}=13$\,TeV simulated with HERWIG (soft tune) are used to train the BDT. Within uncertainties, the $\langle \nmpi^{\rm reg.} \rangle$ values are consistent with the expected ones, suggesting that our approach is robust against hadronization effects.  

The classification of events in terms of \nmpi requires the study of the \nmpi distribution. The regression in this case is performed considering the full set of input variables described in section 2: spherocity, average \pt and forward multiplicity. Figure~\ref{fig:1b} shows the distribution of the self-normalized \nmpi. The comparison of the true self-normalized \nmpi distribution to those obtained from regression indicates that the high \nmpi tail can be better described if simulations with a flat distribution of \nmpi is used to train the BDT. Therefore, the event classifier sensitive to \nmpi ($\nmpi^{\rm reg}$ ) is trained using simulations which consider a flat distribution of \nmpi. 

Figure~\ref{fig:2} shows the behavior of particle production as a function of \nmpi (left),  $\nmpi^{\rm reg}$ (middle) and charged-particle multiplicity at mid-pseudorapidity, ${\rm d}\nch/{\rm d}\eta$ (right), in pp collisions at $\sqrt{s}=2.76$\,TeV. The results are qualitatively similar at other center-of-mass energies including 13\,TeV. Here, the results for 2.76\,TeV are shown to illustrate that albeit the BDT were trained for 13\,TeV, its discrimination power holds even for lower energies. The study includes the proton-to-pion ratio as a function of \pt, as well as a quantity called   $R_{\rm pp}$, which is motivated by the nuclear modification factor used to quantify parton energy loss effects in heavy-ion collisions~\cite{Acharya:2019yoi}. In the case of the analysis as a function of \nmpi, $R_{\rm pp}$ is defined as follows:

\begin{figure}[b]
\includegraphics[width=0.5\textwidth]{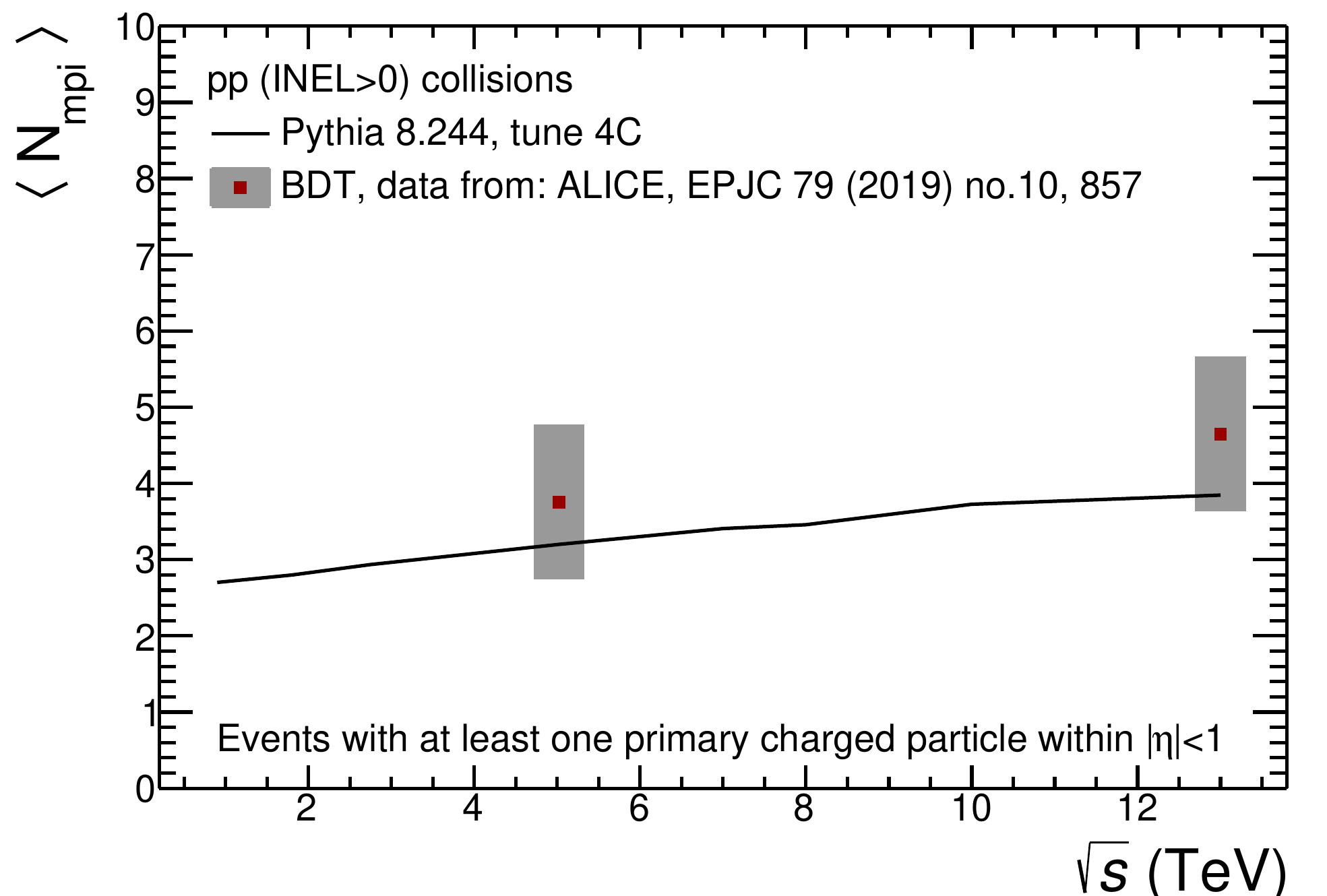}
\caption{\label{fig:3} Average number of MPI as a function of $\sqrt{s}$. Results from PYTHIA~8.244 (solid line) are compared to the estimated $\langle \nmpi \rangle$ (markers) obtained from the application of the trained BDT to the existing ALICE data~\cite{Acharya:2019mzb}.}
\end{figure}

\begin{equation}
R_{\rm pp}=\frac{ {\rm d}^{2}N_{\pi}^{\rm mpi}/(\langle {\rm N}_{\rm mpi}\rangle {\rm d}y {\rm d}\pt) }{ {\rm d}^{2}N_{\pi}^{\rm MB}/(\langle {\rm N}_{\rm mpi,\,MB}\rangle {\rm d}y {\rm d}\pt) }
\label{eq:2}
\end{equation}
where,  $ {\rm d}^{2}N_{\pi}^{\rm MB}/{\rm d}y {\rm d}\pt$ and $ {\rm d}^{2}N_{\pi}^{\rm mpi}/{\rm d}y {\rm d}\pt$ are the charged-pion yields for the MB sample and for a particular subsample defined by its MPI activity, respectively. Given the requirement for spherocity calculation, the MB sample corresponds to events with more than two primary charged particles within $|\eta|<0.8$ and $\pt>0.15$\,GeV/$c$, however, the conclusion remains the same for the most inclusive sample. In analogy to the normalization to the number of binary collisions which enters in the definition of $R_{\rm AA}$, the yields are normalized to their corresponding $\langle {\rm N}_{\rm mpi}\rangle$. Therefore, $R_{\rm pp}$ is the pion yield per semi-hard partonic scattering in high-\nmpi (or low-\nmpi) pp collisions normalized to that for MB pp collisions.  In the absence of QGP effects or selection bias, $R_{\rm pp}$ is expected to be unity at high \pt.  For the event selection based on $\nmpi^{\rm reg}$ (${\rm d}\nch/{\rm d}\eta$), $\langle {\rm N}_{\rm mpi}^{\rm reg}\rangle$ ($\langle {\rm d}N_{\rm ch}/{\rm d}\eta \rangle$) is used in Eq.~\ref{eq:2} instead $\langle {\rm N}_{\rm mpi}\rangle$.

Regarding the analysis as a function of \nmpi (top left panel), while $R_{\rm pp}$ is \nmpi independent and close to unity~\footnote{Actually, $R_{\rm pp}$ is slightly below one as the reference corresponds to events with multiplicity above a given threshold. This introduces a small bias in the reference (denominator) towards hard events.} at high \pt ($\pt>8$\,GeV/$c$), $R_{\rm pp}$  develops a bump at intermediate \pt (1-8\,GeV/$c$). The former effect is consistent with a binary parton-parton scaling which holds given the absence of any parton-energy loss mechanism in PYTHIA~\cite{Ortiz:2020dph}. Regarding the behavior at intermediate \pt, the bump is attributed to color reconnection~\cite{Ortiz:2018vgc,Ortiz:2013yxa}, which is known to mimic collective effects. This is consistent with our results for simulations without CR, which give $R_{\rm pp}$ nearly flat and independent of \nmpi. The bump at intermediate \pt resembles the Cronin effect~\cite{Cronin:1974zm} observed in p--Pb collisions~\cite{Acharya:2018qsh}. Albeit the effect is rather large ($\approx40$\%), it is worth mentioning that, given the limitations of the multiplicity estimators used in the experiments~\cite{Acharya:2019mzb}, the bump has not been observed in pp data~\cite{Acharya:2020zji}. 

We want to highlight the fact that using regression, one can reduce the selection bias and increase the sensitivity to \nmpi. The top middle panel of Fig.~\ref{fig:2} shows the results as a function of $\nmpi^{\rm reg}$, which qualitatively (and sometimes quantitatively) recovers the main characteristics of the \nmpi dependence. The plot also includes the uncertainty related to event selection, which is shown as boxes around one. It has been derived from the average deviation of \nmpi with respect to $\nmpi^{\rm reg}$, it is around 30\% at $\nmpi^{\rm reg}=4$ and it is reduced to 17\% at higher $\nmpi^{\rm reg}=15$. It is worth noticing that the effects discussed above are larger than such uncertainties. The implementation of this event selection in pp and p--Pb LHC data would definitely provide valuable information on the production mechanisms, as well as it will help in understanding the similarities with larger systems like those created in A--A collisions in a better way.

The top right-hand-side panel of  Fig.~\ref{fig:2} shows the results when the analysis is performed as a function the mid-pseudorapidity estimator.  In this case, the bahaviors discussed before for intermediate and high \pt  are not observed. Actually, what we observe is the effect of autocorrelations given that the event activity (multiplicity) and the \pt distributions are both determined within the same pseudorapidity interval.

Last but not least, we point out that the size of the bump in $R_{\rm pp}$ is hadron mass dependent, whose behavior resembles the features of the $R_{\rm p-Pb}$ for identified hadrons~\cite{Adam:2016dau}. To illustrate the hadron mass dependence as a function of the event activity, the bottom panel of Fig.~\ref{fig:2} shows the proton-to-pion ratio as a function of \pt for the event classes described above. As reported in Ref.~\cite{Ortiz:2013yxa}, the particle ratio gets depleted (enhanced) at low (intermediate) \pt with increasing \nmpi. A similar feature is also observed when the analysis is performed as a function of $\nmpi^{\rm reg}$. In simulations without color reconnection, the particle ratios are independent of \nmpi and $\nmpi^{\rm reg}$ within a few percents. The effect is not observed when the analysis is performed as a function of the charged-particle multiplicity at mid-pseudorapidity. 

Finally, using the existing ALICE data on \pt spectra as a function of mid-pseudorapidity multiplicity,  the average number of MPI was estimated using the trained BDT.  Figure~\ref{fig:3} shows the number of MPI values obtained from regression along with PYTHIA~8.244 calculations. In our approach, the average number of MPI in (${\rm INEL}>0$) pp collisions at $\sqrt{s}=5.02$ and 13\,TeV are found to be 3.76$\pm1.01$ and 4.65$\pm1.01$, respectively. The ${\rm INEL}>0$ class defined by ALICE corresponds to pp collisions with at least one primary charged particle within $|\eta|<1$. 

To test the robustness of this result, the BDT were applied to HERWIG~7 simulations~\cite{Bellm:2019zci}. The average \nmpi obtained from regression was $4.3\pm1$ and $1.8\pm1$ for simulations with and without MPI, respectively. The fact that $\langle \nmpi^{\rm reg.} \rangle$ for MPI=off is slightly above the expected value (one), suggests that the results have a small model dependence which is covered by the systematic uncertainties.  

In summary, we propose the use of multivariate techniques in order to build more robust event classifiers for a better understanding of the similarities observed in different collision systems. The proposed event classifier can be used to test the MPI model in bigger systems like p--A and A--A collisions. Also, it can be used to refine the jet quenching searches in small systems~\cite{Ortiz:2020dph}.

\section{Conclusions}
In this work, we have proposed a new event classifier to analyse the pp data at the LHC. We have shown that using input variables like charged-particle multiplicity, average transverse momentum and transverse spherocity, one can build an event classifier sensitive to the number of partonic interactions (\nmpi). The target variable $\nmpi^{\rm reg}$ was used to build $R_{\rm pp}$, which is analogous to the nuclear modification factor used in A--A collisions to study the parton-energy loss effects. Within uncertainties, this quantity is independent of $\nmpi^{\rm reg}$ and close to unity at high \pt ($>8$\,GeV/$c$). Moreover, at intermediate \pt (1-8\,GeV/$c$) a bump is observed in events with large event activity. The effect is attributed to multi-parton interactions and color reconnection, and has not been observed in data. The trained methods were also applied to the available ALICE data on \pt spectra as a function of multiplicity. In our approach, the average number of MPI in (${\rm INEL}>0$) pp collisions at $\sqrt{s}=5.02$ and 13\,TeV are 3.76$\pm1.01$ and 4.65$\pm1.01$, respectively. 
\begin{acknowledgments}

\end{acknowledgments}
We acknowledge the technical support of Luciano Diaz and Eduardo Murrieta for the maintenance and operation of the computing farm at ICN-UNAM. Support for this work has been received from CONACyT under the Grant No. A1-S-22917 and from UNAM-PAPIIT under Project No. IN102118. S. T. acknowledges the postdoctoral fellowship of DGAPA UNAM.

\nocite{*}

\bibliography{biblio}

\end{document}